# P2P DOMAIN CLASSIFICATION USING DECISION TREE

Anis ISMAIL, Aziz BARBAR

American University of Science & Technology - AUST
Alfred Naccash Avenue - Ashrafieh
Beirut, Lebanon
anismaiil@ul.edu.lb
abarbar@aust.edu.lb

## ABSTRACT

*The increasing interest in Peer-to-Peer systems (such as Gnutella) has inspired many research activities in this area. Although many demonstrations have been performed that show that the performance of a Peer-to-Peer system is highly dependent on the underlying network characteristics, much of the evaluation of Peer-to-Peer proposals has used simplified models that fail to include a detailed model of the underlying network. This can be largely attributed to the complexity in experimenting with a scalable Peer-to-Peer system simulator built on top of a scalable network simulator. A major problem of unstructured P2P systems is their heavy network traffic. In Peer-to-Peer context, a challenging problem is how to find the appropriate peer to deal with a given query without overly consuming bandwidth? Different methods proposed routing strategies of queries taking into account the P2P network at hand. This paper considers an unstructured P2P system based on an organization of peers around Super-Peers that are connected to Super-Super-Peer according to their semantic domains; in addition to integrating Decision Trees in P2P architectures to produce Query-Suitable Super-Peers, representing a community of peers where one among them is able to answer the given query. By analyzing the queries log file, a predictive model that avoids flooding queries in the P2P network is constructed after predicting the appropriate Super-Peer, and hence the peer to answer the query. A challenging problem in a schema-based Peer-to-Peer (P2P) system is how to locate peers that are relevant to a given query. In this paper, architecture, based on (Super-)Peers is proposed, focusing on query routing. The approach to be implemented, groups together (Super-)Peers that have similar interests for an efficient query routing method. In such groups, called Super-Super-Peers (SSP), Super-Peers submit queries that are often processed by members of this group. A SSP is a specific Super-Peer which contains knowledge about: 1. its Super-Peers and 2. The other SSP. Knowledge is extracted by using data mining techniques (e.g. Decision Tree algorithms) starting from queries of peers that transit on the network. The advantage of this distributed knowledge is that, it avoids making semantic mapping between heterogeneous data sources owned by (Super-)Peers, each time the system decides to route query to other (Super-) Peers. The set of SSP improves the robustness in queries routing mechanism, and the scalability in P2P Network. Compared with a baseline approach,the proposal architecture shows the effect of the data mining with better performance in respect to response time and precision.*

## 1. INTRODUCTION

The Peer-to-Peer computing paradigm is being viewed as a novel approach for people to share resources such as files and computing cycles, or to support collaborative tasks. The popularity of Peer-to-Peer systems in the last couple of years illustrates how the Internet is gradually shifting toward a distributed system that supports more than unique client-server application. Peer-to-Peer (P2P) systems are distributed systems in which nodes of equal roles and capabilities exchange information and services directly with each other. In recent years, P2P has emerged as a popular way to share huge volumes of data. The key to the usability of a data-sharing P2P system, and one of the most challenging design aspects, is efficient techniques for search, route queries and retrieval of data. The major problem in such networks is query routing, i.e. deciding to which other (Super-)Peers the query has to be sent for high efficiency and effectiveness. The tradition P2P systems offer support for richer queries than just search by identifier, such as keyword search with regular expressions. Search techniques for these systems must therefore operate under a different set of constraints than that of the techniques developed for persistent storage utilities.

However, such systems that broadcast all queries to all peers suffer from limited efficiency and scalability. In hybrid P2P systems [1][2], composed of (Super-)Peers, when a peer submits a query, this peer becomes the source of this query. Then the query is transmitted to its Super-Peer (SP). The routing policy in use quickly determines the relevant neighbors (SP), based on semantic mappings between schemas of (Super-)Peers; and, to which neighbours the query is to be sent. When an SP receives a query, it processes the query over its local collection of data sources of different peers. If any results are found, the SP sends a single response message back to the query source. Another important aspect of the user experience is the long time duration the user must wait for the results to arrive. This is due mostly to the mediation process which remains difficult to realize in such a context when the number of (Super-)Peers increases. Response times tend to be slow in hybrid P2P networks, since the query travel through several SP in the network; and where the SP is forced to look for connections (i.e. mappings) in order to route the query. Satisfaction time is simply the time that has elapsed from when the query is first submitted by the user, to when the user receives the overall results.

Data mining has recently become very popular due to the emergence of vast quantities of data. In this paper, a practical issue about data mining in P2P network is discussed. The motivations behind P2P data mining include the optimal usage of available computational resources, privacy and dependability upon eliminating critical points of service.

In this paper, the effect of data mining in P2P query routing is presented. The proposed method focuses on how the query is routed to relevant peers with minimum query processing at SP level in order to improve answering time of the queries. The important advantage of this approach is scalability.

The proposed approach consists of grouping together (Super-)Peers that have similar themes for an efficient query routing method. Each obtained group, called Super-Super-Peers (SSP), contains domains, and is composed of Super-Peers (the responsible domains) and their corresponding peers (the members) that submit queries that are often processed by members of this group. Each SSP operates with an index that is obtained by applying Decision Tree algorithms at the same time keeping track of where contents concerning a query are located. When an SSP receives a query from a Super-Peer (in his group), it directly consults its index (without making any mappings) in order to determine 1. In this group all Super-Peers (or domains) that are able to answer this query and 2. In other groups (i.e. other SSP) all Super-Peers which are relevant to this query are found.

This paper further discusses the said topics in details in the following sections, starting with a short presentation of related work, then in section 3 presents, in brief, the principal concepts of P2P networks, showing the context of in which the proposed work was executed. Section 4, presents the baseline algorithm of queries routing in hybrid P2P systems; and Section 5, introduces the Super-Super-Peer (SSP) network with Decision Tree. Section 6 presents the semantic routing of queries algorithm, while the suggested simulator used to evaluate the implemented approach is presented in Section 7. Section 8 shows some experiments and evaluations, to end in Section 9 with the conclusions.

## 2. RELATED WORK

P2P networks are quickly emerging as large-scale systems for information sharing. Through networks such as Kazaa, e-Mule, BitTorrents, consumers can readily share vast amounts of information. While initially the consumer’s interest in P2P networks was focused on the value of the data, more recent research such as P2P web community formation argues that the consumers will greatly benefit from the knowledge locked in the data presented by Liu et. al and Bhaduri et al. [3][4].

Efficient query routing in P2P systems has already been discussed in the literature [16][17]. Semantic query routing techniques are required to improve effectiveness and scalability of search processes for resource sharing in P2P systems. The unstructured P2P systems typically employ flooding and random walk to locate data, which results in much network traffic.

Query routing in a Peer-to-Peer network is the process by which the query is routed to a number of relevant peers, and, consequently, it is not broadcasted to the whole network. The problem of query routing concerns the discovery of relevant peers to such query after which peers are considered as relevant are denoted. Accordingly, first, the criteria by which whether a peer is relevant or not is defined. For example, in some P2P systems, relevant peers are the ones that match exactly all the query predicates. Secondly, the strategy on which routing will be based (e.g. based on routing indices) and all the required routing steps are defined.

In Peer-to-Peer systems, the network topology and the category of P2P determine, to a large extent, the applied routing strategy. Hence, before describing a routing algorithm, it is imperative to look at the characteristics of the Peer-to-Peer network that is to be applied. An efficient query routing aims to limit consuming network bandwidth by reducing messages across the network, and reducing the total query processing cost through minimizing the number of peers that contribute to the query's results. Finally, routing in P2P networks is crucial for the scalability of the network. In the next subsections the dominant approaches of query routing and their applied peer-to- peer environment are described.

Wolfgang Nejdl et. al [5][6][7] presented the routing approach based on routing indices. This approach has been suggested and adapted under various scenarios. It is built upon an RDF-based Peer-to-Peer network. Queries and answers to queries are represented using RDF metadata, and which can be used together with the RDF metadata to describe the content of peers to build explicit routing indices; thus, facilitating the more sophisticated routing approaches. Queries can then be distributed relying on these routing indices which contain metadata information plus appropriate pointers to other (neighboring) peers indicating the direction where specific metadata (schemas) are used. These routing indices do not rely on a single schema, but can contain information about arbitrary schemas used in the network. The recommended approach is based on routing distributed indexes in order to find the Super-Peer with minimum query processing, which is the strength of this recommended approach over the previous one.

The advanced technique presented by Löser and his teams [8][9] is also applied for Super-Peer Schema-Based Peer-to-Peer networks. Based on predefined policies, a fully decentralized broadcast and matching approach distribute the peers automatically to Super-Peers. The basic idea here is that the Super-Peer establishes and maintains a specific Semantic Overlay Cluster (SOC). SOCs define peer clusters according to the metadata description of peers and their contents. Similar to the creation of views in database systems, the Semantic Overlay Clusters are defined by human experts. They act as virtual, abstract, independent views of selected peers in a Schema-Based P2P system. As for the proposed approach, the architecture is built by regrouping the Super-Peers according to their interest, at the same time integrating in each group an index (Decision Tree) to find the relevant Super-Peer and other groups in an intelligent way. Another approach for query routing is presented by Tempich et. al in a study entitled "REMINDIN': Semantic query routing in Peer-to-Peer networks based on social metaphors", [10], defines a method for query routing called REMINDIN' (Routing Enabled by Memorizing INformation about DIstributed INformation). This routing method allows peers to observe which queries are successfully answered by others, memorizes this observation and subsequently uses this information in order to select peers to forward requests to. The basic steps of REMINDIN' routing method are: 1) selecting (at most) two peers from a set of known peers based on a given triple query; hence, avoiding network flooding; 2) memorizing this observation; 3) forwarding the query to the selected peers; and, 4) assessing and retaining knowledge about which peer has answered which queries successfully.

Raahemi, Hayajneh and Rabinovitch [11] present a new approach using data-mining technique, in particular a Decision Tree, to classify Peer-to-Peer (P2P) traffic in IP networks by capturing Internet traffic at a main gateway router, by performed preprocessing on the data, selected the most significant attributes, and prepared a training-data set to which the decision-tree algorithm was applied. They built several models using a combination of various attribute sets for different ratios of P2P to non-P2P traffic in the training data. They observed that the accuracy of the model increases significantly when they include the attributes "Src IP addr" and "Dst IP addr" in building the model. By detecting communities of peers, we achieved classification accuracy of higher than 98%. However, our approach uses data-mining (Decision Tree) to classify the Super-Peers (communities). By detecting communities (domains) of peers, we achieved classification accuracy of higher than 99%.

Roussopoulos, Baker, Rosenthal, Giuli, Maniatis and Mogul [12] present a heuristic Decision Tree that designers can use to judge how suitable a P2P solution might be for a particular problem. It is based on characteristics of a wide range of P2P systems gleaned from the literature. This includes budget, resource relevance, trust, rate of system change, and criticality. Bhaduri, Wolff, Giannella and Kargupta [13] propose a P2P Decision Tree induction algorithm in which every peer learns and maintains the correct Decision Tree compared to a centralized scenario. This algorithm is completely decentralized, asynchronous, and adapts smoothly to changes in the data and the network. This technique offers a scalable and robust distributed algorithm for Decision Tree induction in large Peer-to-Peer (P2P) environments. Computing a Decision Tree in such large distributed systems using standard centralized algorithms can be very communication-expensive and impractical because of the synchronization requirements.

Data mining over multiple data sources has emerged as an important practical problem with applications in different areas such as data streams, data-warehouses, and bioinformatics. Although the data sources are willing to run data mining algorithms in these cases, they do not want to reveal any extra information about their data to other sources due to legal or competitive concerns. One possible solution to this problem is to use cryptographic methods. However, the computation and communication complexity of such solutions render them impractical when a large number of data sources are involved. F. Emekci, O.D. Sahin, D. Agrawal, and A. El Abbadi [14] consider a scenario where multiple data sources are willing to run data mining algorithms over the union of their data as long as each data source is guaranteed that its information that does not pertain to another data source is not revealed. Emekci et. al focus on the classification problem in particular, and present an efficient algorithm for building a Decision Tree over an arbitrary number of distributed sources in a privacy preserving manner using the ID3 algorithm.

Medview [15] was designed earlier to support the learning process, and provide a computerized teaching aid in oral medicine and oral pathology. MEduWeb is a web-based educational tool that allows students to search the database and generate exercises with pictures of real patients [15]. MEduWeb uses the MedView database containing several thousand patient examinations; and on which Khan, Anwer, Torgersson and Falkman use Data mining technique (Decision Trees) [15]. The authors explored the possibilities of using Data mining technique (Decision Trees) on the P2P database, and have performed a series of experiments.

# 3. BACKGROUND

## 3.1. Basic notions

A Peer is an autonomous entity with a capacity of storage and data processing. In a computer network, a Peer may act as a client or as a server. A P2P is a set of autonomous and self-organized peers (P), connected together through a computer network. The purpose of a P2P network is the sharing of resources (files, databases) distributed on peers by avoiding the appearance of a peer as a central server in this network. We note: P2P = (P, U), P is the set of peers and U represents links (overlay connections) between two peers $P_i$ and $P_j$, $U \subseteq P \times P$. The hybrid P2P (P2Ph) (See Figure 1) network that we consider in this paper includes sets of peers (P) and Super-Peers (SP). We note : P2Ph = ($P \cup SP$, K), where P is the set of peers, SP is the set of Super-Peers and K is the set of overlay links expressed under the format of pairs : ($P_i$, $SP_j$ ) or ($SP_j$ ,$SP_k$) which respectively link a Peer $P_i$ to a Super-Peer $SP_j$ or a Super-Peer $SP_j$ to one or several Super-Peers $SP_k$.

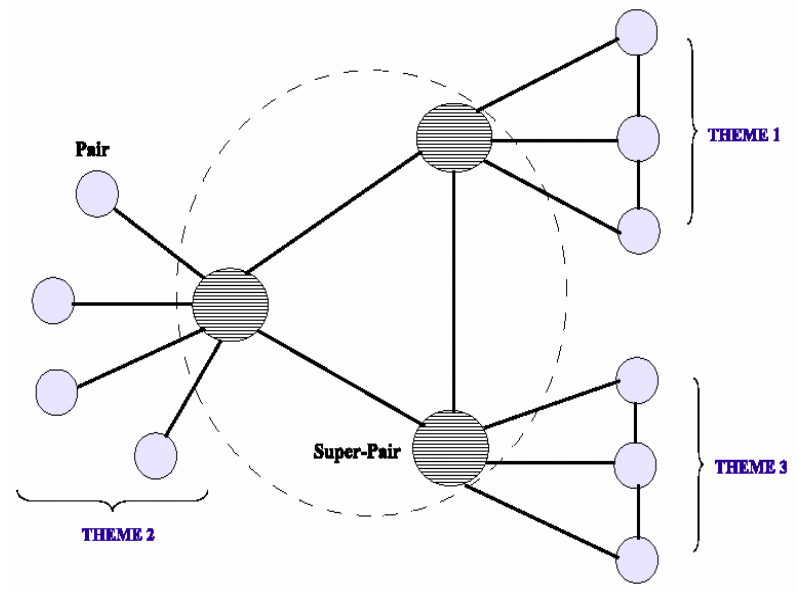

Figure 1. Hybrid network (P2Ph)

A PDMS (Peer Data Management System) combines P2P systems and databases systems. The PDMS that is considered in this paper is a scale hybrid system $P2P^h$. Each peer is supposed to hold a database (or an XML document, etc.) with a data schema. Each Super-Peer provides a theme (a semantic domain, a subject, or an idea) representing special interest to a group of peers. The themes are not necessarily separated; they are described by Super-Peers, with the three following manufacturers:

– A concept is a collection of individuals that constitute the entities of the modeled domain. The concepts can be compared to the notion of class (i.e. object model) or type of entity in the conceptual models (i.e. Entity/Relationship).

– A role is a binary relationship between concepts. Roles are used to specify properties of instances and are compared to the notion of attributes in the conceptual models. A role is viewed as a function that links a concept (called domain) to another concept (known as co-domain).

– Specialization (IsA) starts from a specific concept to a more general concept. It is transitive and asymmetric**,** and defines a hierarchy between concepts that it connects.

We note R as the set of relations reduced in this paper to two relations that are {Role; IsA} and PDMS={PS $\cup$ SP, T, D , K} where PS represents all the peers of the network with their data schemas S={$S_1$, …., $S_p$}. A peer is connected to the network with only one data schema. K is the set of overlay links between (Super-)Peers. Each peer P $\in$ PS is doted of a Data Management System (denoted DMS), and is able to manage their data. TJ={T1,…., Tk} represents the interest themes published by Super-Peers SP through the network. In the proposed approach, each Super-Peer publishes only one theme and peers express their interests in one or several theme(s) in T. The themes are not disjoints: two Super-Peers can publish the same concepts or roles with distinct structures and/or don't use the same vocabulary. DJ = {D1, …., Dk} describes the themes in the set of T: $D_j$ describes the theme $T_j$ specifying the set of concepts and their relationships.

## 3.2. Expertise, Mapping and Domains

At this step, only data models supported by peers are considered. We distinguish the three following data models, the best known: relational, XML and object. An expertise is defined, in our case, as (a part of) the data schema, expressed with one of the three data models cited above, possessed and published by a Peer in order to share its data with other peers. To facilitate the reconciliation, between the data schema of the Peer and the theme described by a Super-Peer, two measures were taken: 1. the expertise of a Peer is expressed with the language of its Super-Peer (i.e. concept, role and IsA); 2. The expertise of a Peer is expressed under the format of couple of elements, satisfying the following condition:

$$\text{EXP}(\text{P}_\text{i}) = \{\theta(S_i;S_j) \in SP \mid (S_i;S_j) \wedge \theta \in R\}$$

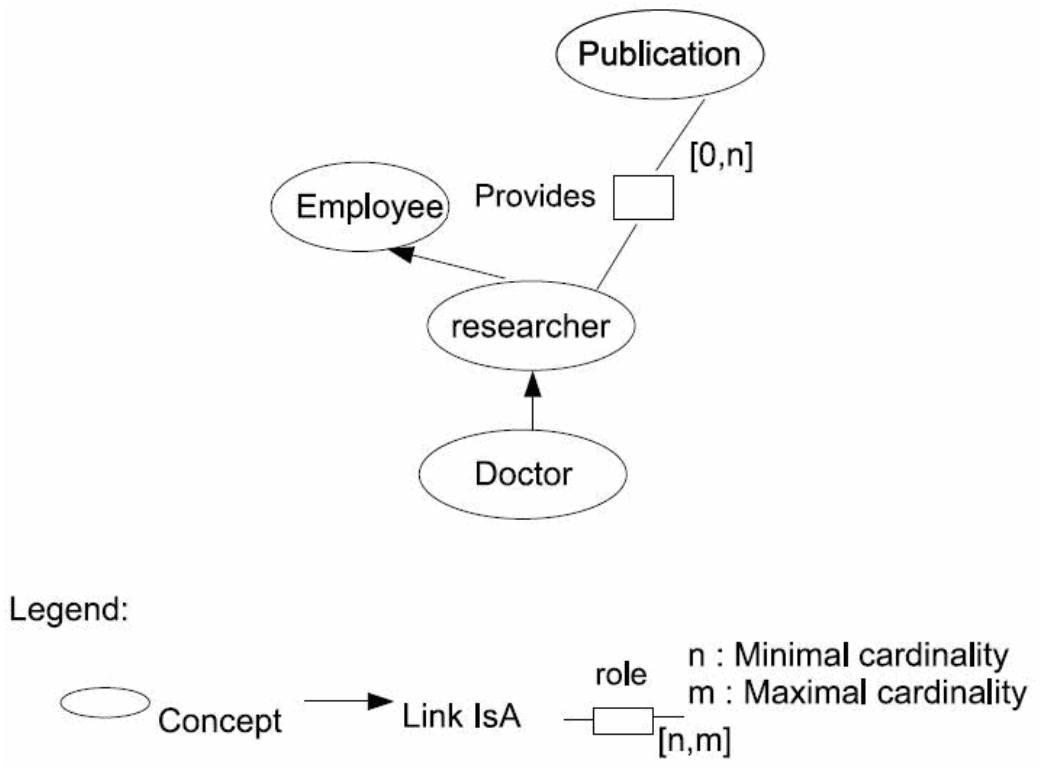

Figure 2. A part of Hospital theme published by SPj

The concepts of this schema of the Peer are: Employee, Publication, Researcher and Doctor. Some links are established between the concepts: Employee and Researcher (see Figure 2). This link expresses that a Researcher is an Employee. The expertise of Pi is given as follows:

EXP (Pi) ={IsA(Researcher;Employee);provides (Researcher; Publication); IsA(Doctor;Researcher)}.

In our context, mapping is an important process in order to share data between peers. Two levels of mapping are distinguished: the first level is to share data between peers; a level which is important while searching for connections between expertise of peers and the description of themes provided by Super-Peers. The second level is to process users' queries, it is important to search for connections between the subject of a query (detailed below) and the expertise of each (Super-)Peers in order to know the group's capacity to response to this query. Let S1, be the expertise of a peer and S2 the theme proposed by the Super-Peer of its domain. The search for correspondence between S1 and S2 is to find for each concept or role in S1 (or S2) a correspondent in S2 (or S1) which is the nearest semantically. We can define the concept of mapping (Map) between schemas as follows:

Map: S1 →S2 Map(es1) = es2 if (1)

Sim(es1; es2) > acceptable-threshold

Where es1: entity of schema S1; es2: entity of schema S2; Sim(es1; es2) is a function, that measure the similarity between two entities es1 et es2, given as follows:

$$\mathrm{Sim}: S_1 x S_2 \rightarrow [0;1] \quad (2)$$

We distinguish two particular cases: Sim(es1; es2) = 1 describes two similar entities ;

Sim(es1; es2) = 0 describes two distinct entities.

We introduce the two concepts, Semantic Intra-Domain and Semantic Inter-Domain. A Semantic Intra-Domain is an interest domain in which mappings between peers, members of this domain, and the Super-Peer, responsible of this Domain, are established. A Semantic Inter-Domain is a set of semantic Intra-Domain in which mappings between Super-Peers of these Domains are established.

We note Semantic Intra-Domain $(SI_a^j C)$ and Semantic Inter-Domains $(SI_a^j C)$ number j (See Figure 3) as follows:

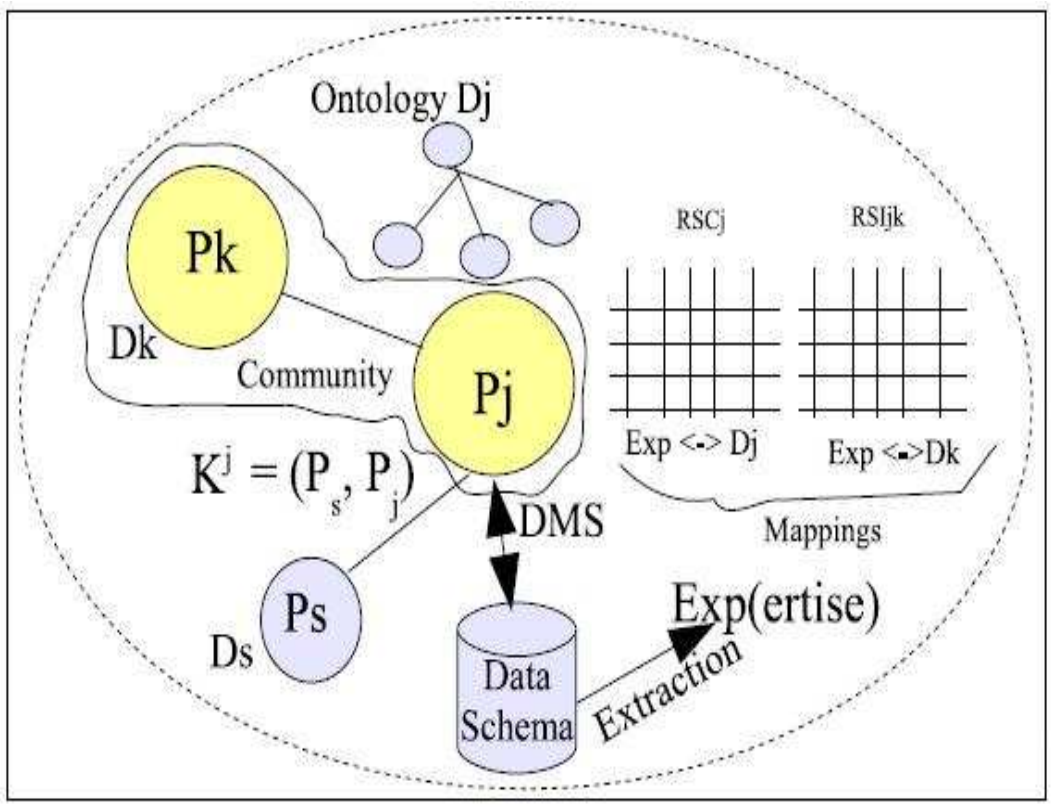

Figure 3. An example of SIeC

$$SI_a^j C = (\mathrm{PS} \cup \mathrm{SP}_{\mathrm{Tj,Dj}}, \mathrm{EXP}(P_s), K^j ; \mathrm{RSC}^j) \quad (3)$$

$$SI_a^j C = (SI_a^j C, RSI^{j-1}, \ldots, RSI^{j-k}) \quad (4)$$

Where k ≠ j, $P_s \subseteq P$ is a subset of peers having the same center of interest Tj; EXP (PS) is the set of expertise of peers interested by this theme and joined to this domain; SPTj, Dj (belong to SP) is the Super-Peer responsible of the domain j which are joined by peers (i.e. a Peer of a domain may request to join several domains if the user thinks that his theme of interest is in the intersection of several domains), Dj represents the description of the theme Tj provided by the Super-Peer. Kj⊆K is the set of overlay links between the Super-Peer SPTj,Dj and the peers connected to it union the set of overlay links between SPTj,Dj and Super-Peers SPTk,Dk, k≠j; RSCj is the semantic Intra-Domain between the Super-Peer SPTj,Dj and the peers inside this domain;RSIj,k is the semantic Inter-Domain concerning the links found between the description of the theme Dj of the Super-Peer SPTj,Dj , with the description Dk of each Super-Peer SPTk,Dk, k≠j). Finally, we introduce a Semantic Overlay Network (SON) represented by the union of all the semantic networks of intra-Domains and inter-Domains. A SON is noted as follows:

$$SON = \bigcup_{j=1}^{|T|} (SI_e^j C) \qquad (5)$$

Where T represents the total number of Super-Peers in the network. The next section presents the query routing algorithm that is our baseline approach.

# 4. SEMANTIC QUERIES ROUTING – BASELINE

## 4.1. Network Configuration

A new Peer Pj advertises its expertise by sending, to its Super-Peer, a domain advertisement $DA_j = (PID;\ E^j_{XP}, T_j; \varepsilon_{acc}; TTL)$ containing the Peer ID denoted PID, the suggested expertise $E^j_{XP}$, the topic area of interest Tj , the minimum semantic similarity value $\varepsilon_{acc}$ required to establish semantic mapping between the suggested expertise $E^j_{XP}$ and the theme of its Super-Peer. When receiving an expertise $E^j_{XP}$, a Super-Peer SPa invokes the semantic matching process to find mappings between its suggested schema and the received expertise.

## 4.2. Baseline approach

A Peer submits its query on its local data schema. This query is sent to its Super-Peer responsible for the domain (see Figure 4). The Super-Peer in its turn names based on the index obtained by the process of mediation (first level), the peers of his domain or the other Super-Peers that are able to treat this query. Each submitted query received by a Super-Peer is processed by searching connections (second level of mappings) between the subject of this query and expertise of peers (of the same domain)**,** or the description of themes of other Super-Peers. In its turn, a Super-Peer from the nearby domain, having received this request, researches among peers (in his domain) that are able to answer this query. The major problem of this approach is the mediation at the two levels cited above: if we take thousands of peers or Super-Peers this approach can not be scaled due to the mappings at both levels. The followings sections describe our approach in order to avoid Super-Peer, when it's too busy to treat all users' queries to process the second level of mapping. This approach improves response times of queries and scalability in P2Ph context by restructuring the network dynamically. To do that, we introduce the concept of Super-Super-Peer (SSP).

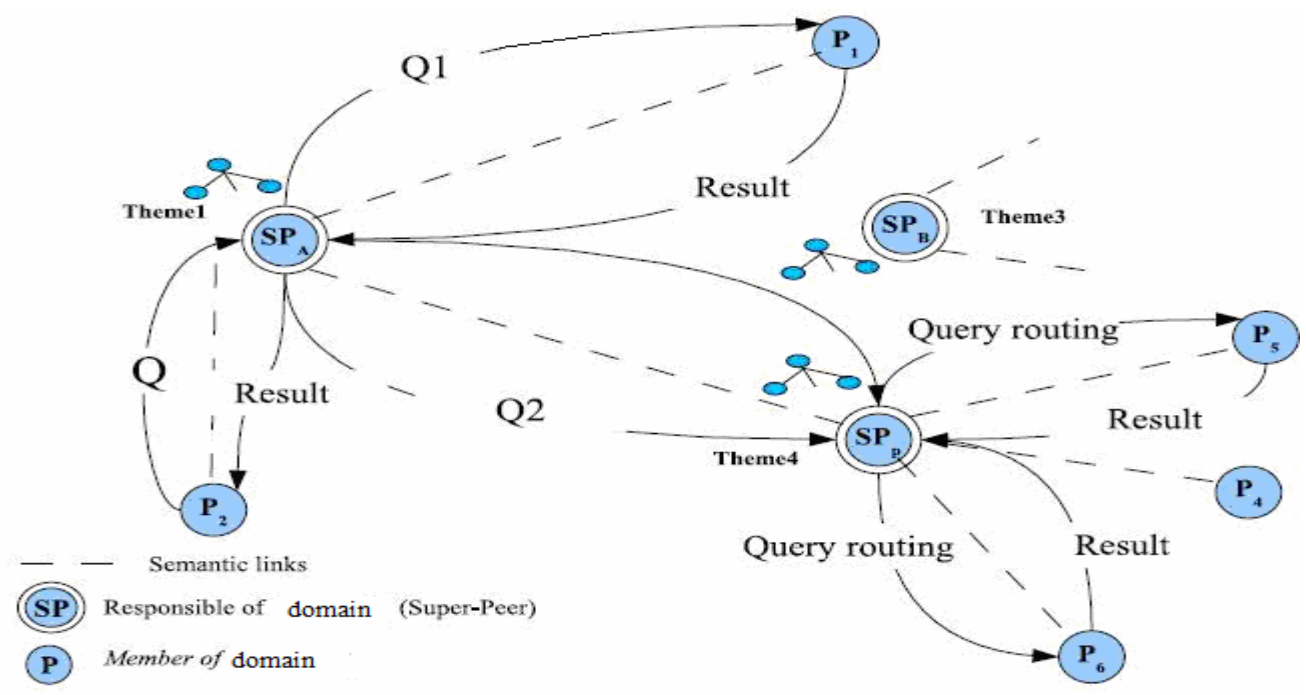

Figure 4. Network configuration and query routing (baseline approach)

## 4.3 Architecture of a Peer

In this section the logical architecture (Figure 5) of a Peer in accordance with the context and topology of the system SenPeer [5] is presented. The role of Peers is to formulate queries and/or respond to queries from remote Peers. The Peers do not perform tasks such as indexing or distributed query processing.

- Data Source: Each Peer has a data model and a data management system to manage its data stored as: a relational database, an XML or RDF. And a query language (SQL, XQuery) related to its data model.
- Wrapper (Adapter): The value of this component is to allow Peers to share their data with other

Peers. This is possible by rewriting the Peer's queries in the common query language SQUEL and vice versa. Wrappers also help to explore the correspondences by transforming the local schema to interchange format schema sGraph.

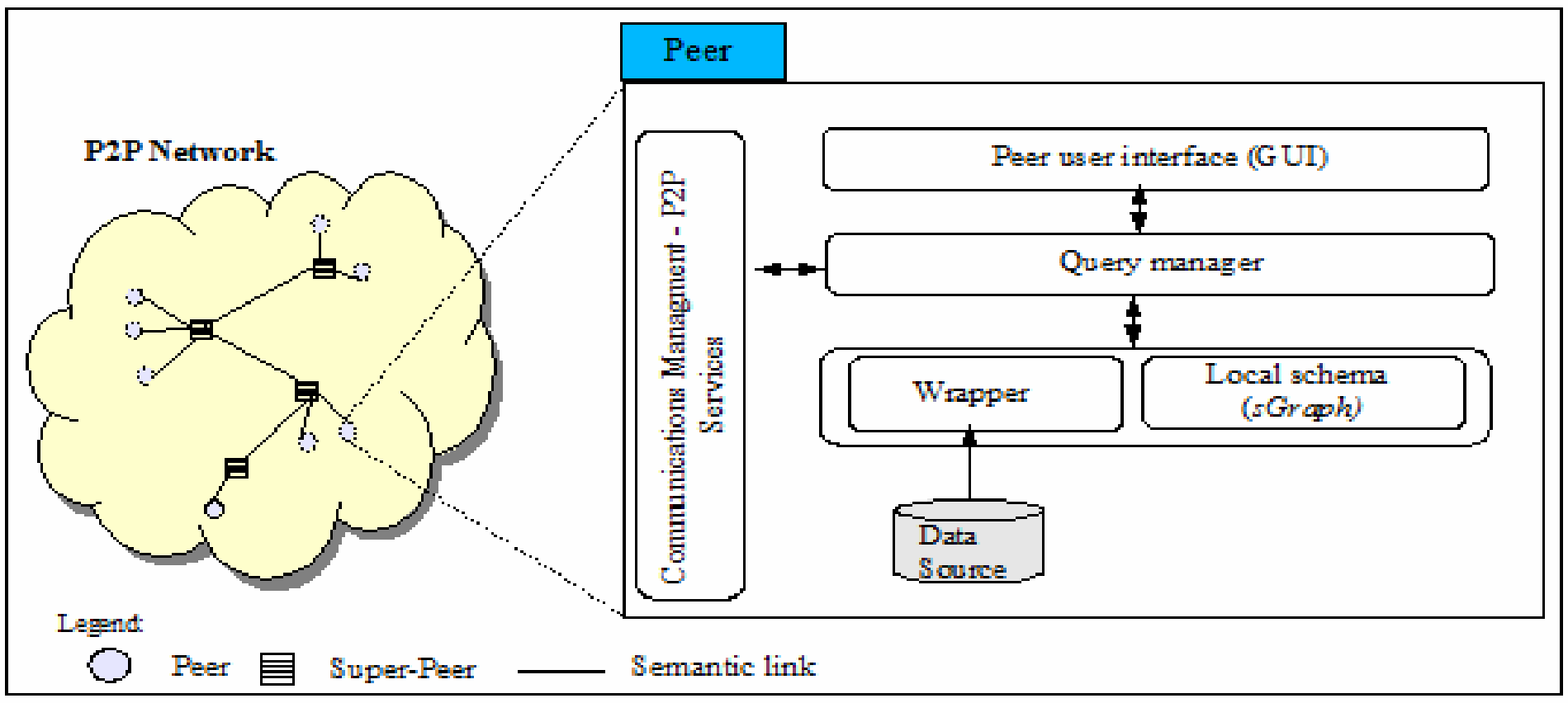

Figure 5. Peer Architecture

- Local semantic network (sGraph): Data published by the Peer is an abstract form of a semantic network (sGraph) with an annotation of the nodes by a set of keywords. This internal model is the semantic representation of the contents of the Peer. It facilitates the exchange Schema among Peers and aims at overcoming the syntactic heterogeneity of Peers to facilitate the discovery of semantic mappings within and/or inter-domain (s).
- Query Manager: This module allows expressing queries local or global. Local queries are executed on the local data source of Peer. Remote queries are routed to the Super-Peer of a Peer for a broadcast.
- GUI: The interface allows a user to make his/her queries and receive responses. We assume that the user is not aware of schemas of remote Peers; thus, formulating his queries on its local schema.
- Communication Manager: Communication among Peers of the system is ensured by the project Open Source Sun's JXTA [JXTA. www.jxta.org.]. JXTA defines a generic network for building a variety of Peer-to-Peer network while remaining independent of platform, programming language (C or Java) systems (Microsoft Windows, Unix), service definitions ( RMI, WSDL) and network protocols (TCP / IP or Bluetooth).

### 4.3 Architecture of a Super-Peer

The Super-Peers can be heterogeneous in terms of computational capacity, bandwidth. The general architecture of a Super-Peer is described in Figure 6.

- Semantic network domain (suggested schema): A Super-Peer has no data to share with other Super-Peers. Each Super-Peer joins the network by suggesting a scheme sGraph that reflects the semantics of the domain which is responsible.
- Manager Matches: This involves managing the semantic links between the internal schema (sGraph) of each Peer and the schema of its Super-Peer.
- Correspondence Matrices: Correspondence matrices store semantic links found by the manager matches. There are two kinds of matrices: Super-Peer/Super-Peer(MSP/SP) containing the correspondences between the Super-Peers responsible for two domains and Super-Peer/Peer (MSP/P) containing the correspondence between a Super-Peer and Peers.
- Domain Index: This component can store information on Peer local area and those on Super-Peers overseeing remote domains that are semantically related. Among these we distinguish information: IP address, bandwidth, ID (Super-)Peer and expertise of these (Super-)Peers. The expertise is stored in two types of tables: Super-Peer/Super-Peer (ESP/SP) for the expertise of the Super-Peer and related Super-Peer/Peer (ESP/P) for the expertise of Peers in his domain. This expertise will be used (later) for efficient routing of queries to the relevant (Super-)Peers.

- Query Manager: The role of this component is to rewrite and route queries to the (Super-)Peers. It also defines the implementation plans and optimizes front to supervise the implementation across the network.
- Communication Module: As for the Peer communication, it is provided by Sun's JXTA [JXTA. www.jxta.org.].

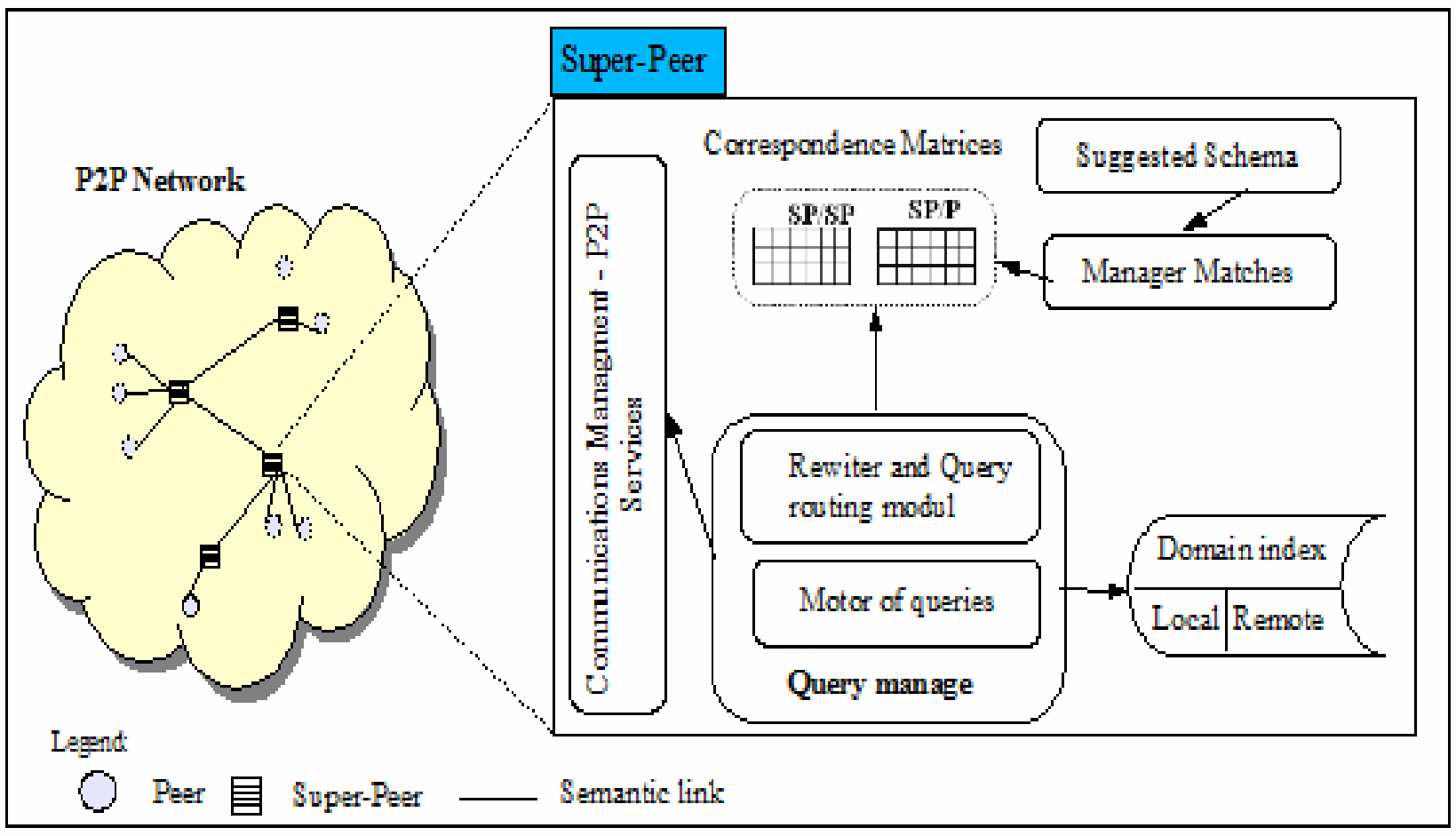

Figure 6. Super-Peer Architecture

# 5. SUPER-SUPER-PEER NETWORK

## 5.1. Topology

A Super-Super-Peer (SSP) network is a semantic sub-network of Overlay Network (SON). The SSP number j is defined as follows:

$$SSP^{j} = \cup_{l=1}^{|M|} (SI_{e}^{l} C) \qquad |M| \leq |T| \qquad (6)$$

Where M is the number of Super-Peer in SSPj; and |M|≤|T| (total number of Super-Peers). $SI_{e}^{l}C$ is the Semantic Inter- domain of the Super-Peer number l.

Two fundamental properties are derived from SSP:

$$SSP^{i} \cup SSP^{j} = SON \text{ , } i \neq j \qquad (7)$$

$$SSP^{i} \cap SSP^{j} = \phi \qquad (8)$$

A Super-Super-Peer is represented physically with a specific Peer. This Peer, representing the Super-Super-Peer number j, is noted as follows:

$SSP^{j} = (PS \cup SP_{TJ,DJ}, EXP(P_{S}), K^{j}, RSC^{J}, RSI^{J}, IND^{j})$ where $PS \subseteq P$ is a subset of peers having very close center of interests denoted T J = {T1,…, Ts}, EXP (PS) is the set of expertise of peers interested by at least one of themes in T J, SPTJ, DJ (belong to SP) is the set of Super-Peers responsible of domains which have very close domain interests, DJ = {D1, …, Ds} represents the description of themes in T J (DJ describes TJ). Kj ⊆K is the set of overlay links between each Super-Peer SPTj, Dj ∈ SPTJ,DJ, and 1) The peers connected to it (within its domain); 2) The other Super-Peers; and, 3) The Super-Super-Peer SSPj itself. RSCJ is the set of semantic Intra-Domain of the Super-Peers ∈ SPTJ, DJ. RSIJ is the set of semantic Inter-Domain for each Super-Peer in SPTJ, DJ. INDj is the index obtained using a Decision Tree algorithm to identify directly the most relevant (Super-)Peers, without going through mappings, to provide good results when a query is submitted by a peer.

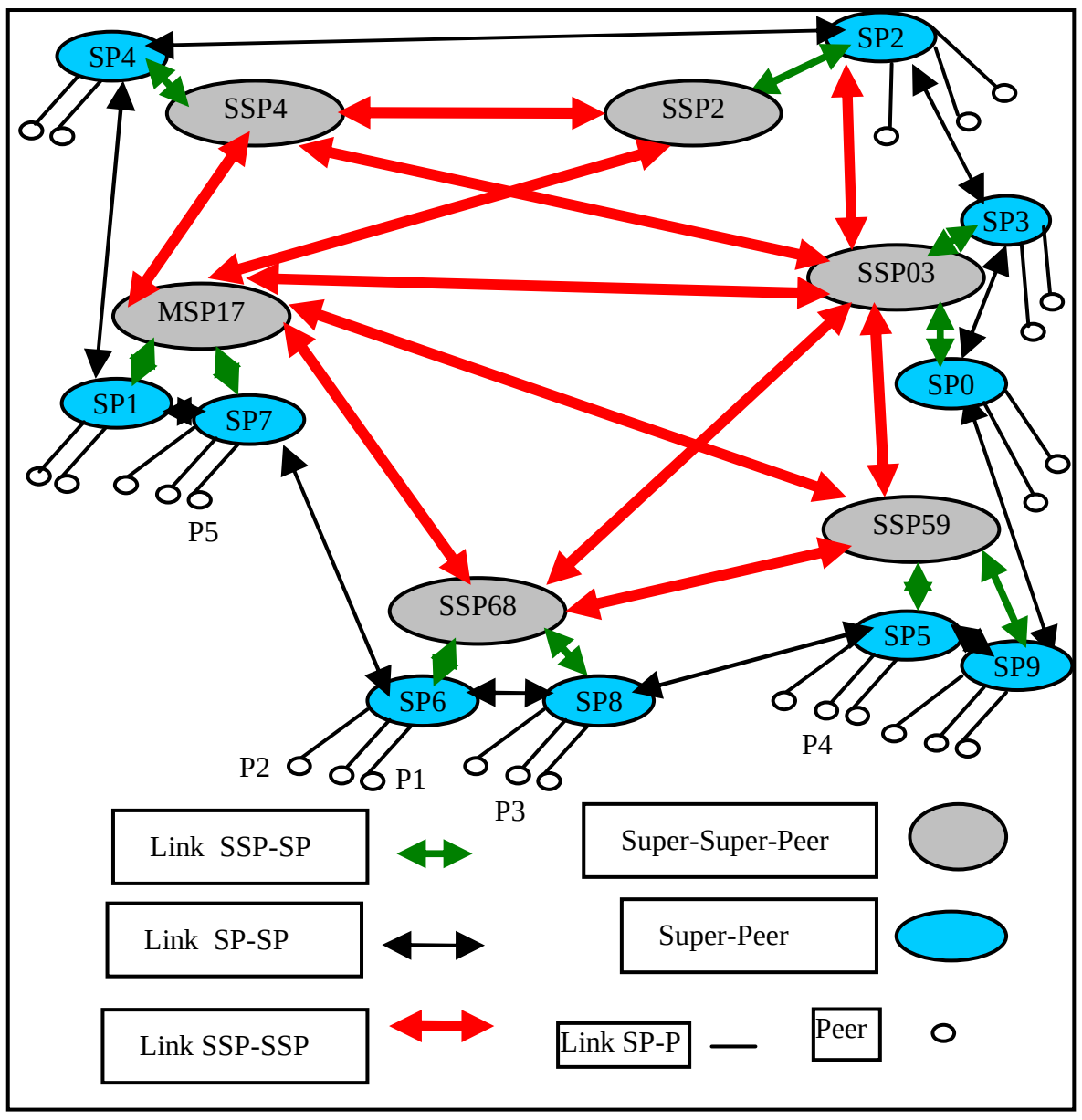

Figure 7. Baseline with Knowledge (DK)

Our proposed System (See Figure 7) is a hybrid P2P system based on an organization of peers around Super-Peers according to their proposed themes, where Super-Peers are connected to a Super-Super-Peer (SSP), the engine that specifies the Super-Peers having peers which may have relevant data to answer queries with minimum query tasks and, by consequence, improve answering time of the queries. The Super-Peer architecture allows the heterogeneity of peers by assigning more responsibility to selected peers. Therefore, certain Peers, called Super-Super-Peers, have an additional computing power and greater bandwidth, resources, performing administrative tasks. They are responsible for routing queries to relevant Super-Peers, not only reducing efforts of compilation of queries**,** but also preventing the spread of queries in the network. In each domain, there is a Super-Peer connected to a Super-Super-Peer where we have an index to identify Super-Peers that are most relevant to provide good results of queries. Otherwise, if the Super-Super-Peer didn't find the relevant Super-Peers form its index for a given query, it returns the query to its parent to work with the baseline to find the answer to this query. We suggest to run our simulations in two configurations, one with the baseline connection between the Super-Peers (Hybrid architecture) (See Figure 7), or else with non connection with the Super-Peers (Distributed knowledge architecture). Figure 8 delineates the effect of the data mining (Decision Tree) in the baseline architecture.

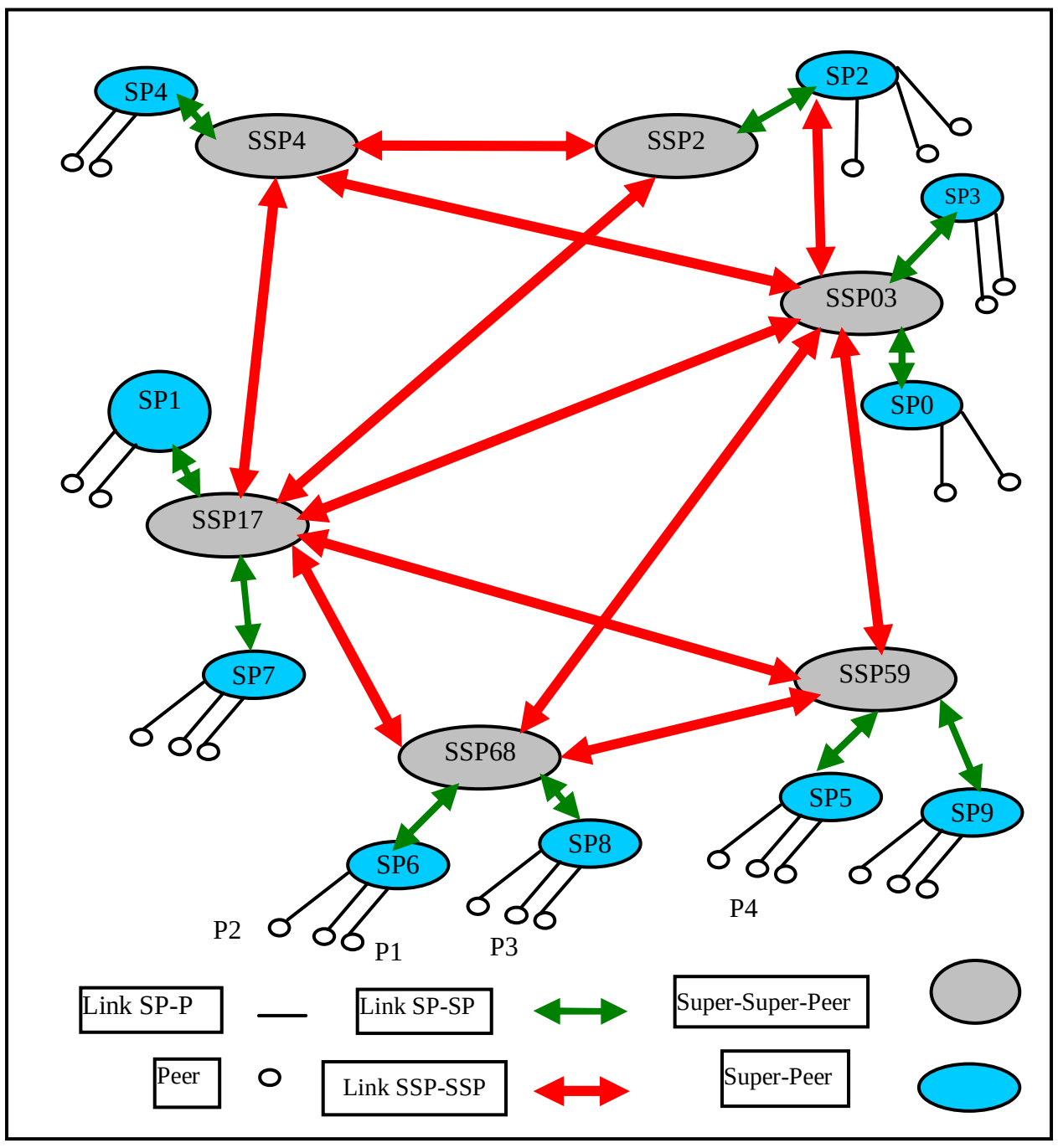

Figure 8. Knowledge only (DK-bis)

The building block (SSP) of the current P2P systems in the architectures (Distributed Knowledge – DK and Hybrid) is the notion of a Super-Peer-group, or a number of nodes (Super-Peer) that participate with each other for a common purpose to minimize the load in the SSP to communicate with other SSP.

## 5.2. SSP Architecture

In this section, we present the logical architecture of a Super-Peer holding knowledge, also known as SSP (Super-Super-Peer). Hereby are the different components related to a SSP: the log file, the discovery of knowledge, a component to predict Super-Peers and relevant request handler. The construction of log file is given by Algorithm 6. The second component is the construction of Decision Trees. Decision Trees are often used for classification and prediction. It is a simple and powerful knowledge representation. The models produced by Decision Trees are represented as tree structures. We used an implementation of an existing algorithm in the WEKA platform to build the Decision Tree from the log file (log). The last component can predict, based on the Decision Tree, all Super-Peers relevant to a query. The query management receives a query from a Super-Peer in the domain-group (SSP) and returns the result of the prediction (the relevant Super-Peers) to the chosen Super-Peer.

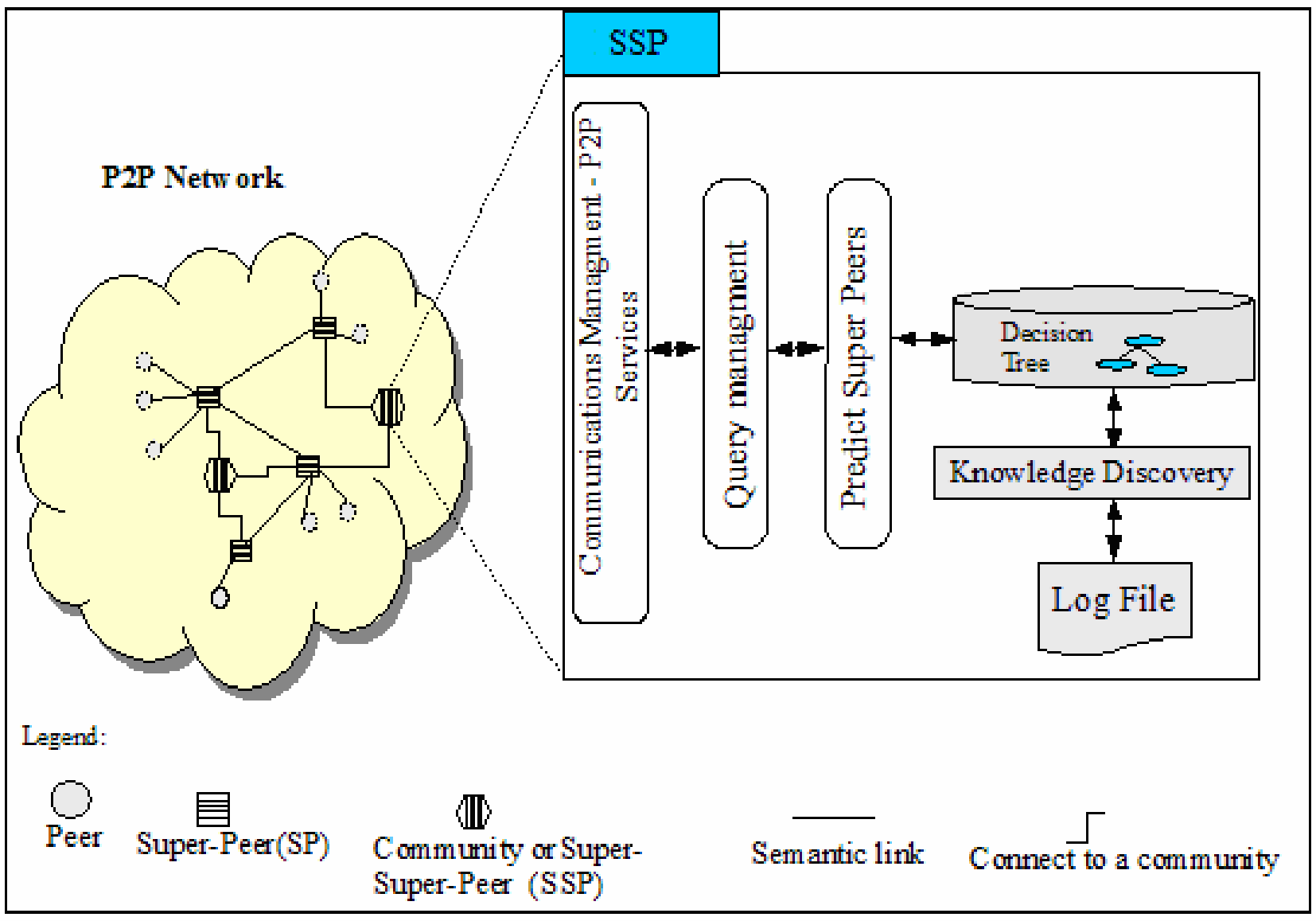

Figure 9. SSP Architecture with all Fields

The general architecture of a Super-Super-Peer is described in Figure 9. Each SSP contains the following components:

- Query Manager: This module differs from that defined by the semantic mediation approach in the fact that the routing application is based on the Decision Tree. This allows the prediction of Super-Peers to which candidates will be forwarded the request to be processed.

- Logfile: It is the file containing the queries processed by a Super-Peer domain. It contains the components of the application and the Super-Peer who has responded to the request. In the case where several Super-Peers respond simultaneously to a query, so many lines will be added to the file Logfile of domain where each line consists of the components of the request followed by a Super-Peer among those who responded to the request. Hence, the number of added lines is in correlation to the number of the Super-Peers who responded to the request.

- Method of construction of knowledge: It is the algorithm for constructing the Decision Tree by analyzing the queries handled by domain-group members and is stored in the Logfile. In our experiments the algorithm J48 WEKA platform to induce the Decision Tree is used.

- **Predication**: This module uses the Decision Tree to predict, for a given query Q, the Super-Peers that are relevant candidates to process the query Q. Contrary to the general case where the tree is used to predict a single value of the class (Super-Peer), in the proposed design we infer all likely values of each class with its own probability. This list of class values is the set of Super-Peers likely to process the request.

## 5.2. Decision Tree

Learning by reasoning is one of the approaches used by Dang [18] to solve problems of extracting knowledge from data. This approach appears in various applications, such as in classification, prediction, by rule extraction from data etc. Several methods have been used for example: neural networks, Decision Trees, Bayesian networks etc.. We are particularly interested in the Decision Tree method to extract knowledge from useful data.

Currently, the algorithms to build Decision Trees (J48, NBTree, ...) allow uniform treatment of almost all types of attributes: numerical, symbolic, fuzzy or probabilistic [18], algorithms also adapted to different problems: the presence of missing data, incorrect, inaccurate, unreliable etc.

The advantages of methods based on Decision Trees have been well detailed in a study conducted by Gay, D. [19]. The predictive model of Decision Tree is easy to analyze, and is applicable to large databases. It also has the advantage of being applicable to digital qualitative data. These characteristics and the functioning of this technique are detailed in the work of CART [20], ID3 [21] and its extension C4.5 [22].

The construction of a generic Decision Tree is presented in Algorithm 1:

**Algorithm 1 : Construction of Decision Trees in general**

Input: DB (X, Y, Z) is a database,
Cl = (CI1, CI2, ..., Clp) The set of classes

Output: Decision Tree DT results

Initialize the empty tree DT;
The current node is the root of the empty tree;
**Do**
**If** the current node is terminal, **then**
assign a class to the current node;
**Otherwise**
Select a test attribute;
Create sub-tree of DT associated with the test's attribute;
**Until** All the leaves are labeled;

For each domain-group Gi, we will associate a Logfile Li, which gathers the queries processed by at least one of its Super-Peers. Then, a knowledge extraction algorithm is triggered to extract the hidden knowledge in the Logfile. This procedure is explained in Algorithm 2:

**Algorithm 2 : LogGroup(Li, Gi), Building file Logfile Li of the domain-group Gi**

*P :* The peer that sent the query
$SP_P$ : Super-Peer of a peer P, which sends the query
$Q_X$ : Query sent by the peer or Super-Peer X
$R_{QP}$ : Answer returned for $Q_P$
Begin
For ((send(P, $Q_P$ ,$SP_P$ ) or send(SPj, $Q_{SPj}$ ,$SP_P$ )) et $SP_P \in$ Gi) Do
Boolean bool = ResearchLocal($SP_P$ ,$Q_P$ ,$R_{QP}$ )
If (bool) Then
Return $R_{QP}$
Updat(**Li,** $SP_P$) // $SP_P$ treated $Q_P$, therefore we add to Li
Else
For $SP_I \in$ neighbor ($SP_P$ ) Do
send($Q_P$ ,$SP_I$ )

bool = process($SP_I$ , $Q_P$ ,$R_{QP}$ )
If (bool) alors
Return $R_{QP}$
Updat(Li, $SP_I$) // $SP_I$ treated $Q_P$, therefore we add to Li
EndIf
EndFor
EndIf
End

The Decision Tree thus predicts a Super-Peer from the components of a query [23]. In general, the Super-Peer returned is the one with the highest probability. We use in this paper a probabilistic inference that can exploit the Decision Tree and return a list of possible Super-Peers, each one with its own probability. Launching of the construction of the Decision Tree is performed periodically according to the number of new queries added to the Logfile Li. Thus, when the number of queries exceeds the threshold θi, then the tree Ti is rebuilt taking in account the new queries of the Logfile of domain-group. The following algorithm specifies the automatic reconstruction driven by changes of the threshold θi

**Algorithm 3 : Reconstruction of the Decision Tree and update of the domain-group**

θi: Threshold of new queries used to reconstruct the tree
Ci: ith the domain-group
Ti: The Decision Tree has Ci domain-group
$\varepsilon_s$: Threshold query to be answered if the Super-Peer is considered inactive

Beginning
If New (Li)> θi then
Ti = cons (Li) // we rebuild the tree Ti
EndIf
For (SP ∈ Ci) do // for all members of the domain-group Ci
If (Score (PS) < $\varepsilon_s$) then / / SP is idle (not active enough)
InviteDeconnect (SP, Ci)
EndIf
EndFor
End

# 6. SEMANTIC QUERY ROUTING ALGORITHM

Our algorithm of semantic query routing is composed of three stages:

– During the first step (the step of baseline approach), the semantic routing algorithm exploits the expertise of (Super-)Peers and the two levels of mappings in order to forward a query q to only relevant Super-Peers. Each Super-Peer in its turn forwards this query to relevant Peers in its domain. The followings sub-steps are necessary in order to process the query:

1. Extract the subject of this query (Sub(Q) (Q of peer P2);

2. Select, by this Super-Peer (SPA), the most relevant peers (P1) for the query and the other Super-Peers (SPP) (by matching the subject of the query to the set of expertise Exp(P2) of peers or to the themes of Super-Peers). The selection is based on a function CAP that measures the capacity of a peer or a Super-Peer on answering a given query;

$$\mathrm{Cap}(P,Q) = \frac{1}{\mathrm{Sub}(Q)} \left( \sum_{s \in \mathrm{Sub}(Q)} \mathop{Max}_{e \in \mathrm{Exp}(P)} S_s(s,e) \right) \qquad (9)$$

3. Once the set of relevant (Super-)Peers has been identified, the Super-Peer sends the query to those promising peers or Super-Peers closest to them by using their ID, IP addresses and the underlying physical network. The advantage of this step is that it permits us, during the second step, to collect information about the queries received by Super-Peers and the relevant super (-peers) selected in order to process it.

– The second step exploits the Hybrid Super-Super-Peers (SSP) network with the baseline approach. This step is very useful when the performance of the system is low. This step runs in four stages:

1. The Super-Peer (SP6) sends the query (Q of P1) directly to its Super-Super-Peer (SSP68);

2. The Super-Super-Peer (SSP68) identifies (without the mapping) the relevant Super-Peers (SP8) that belong to this SSP (SSP68) and other SSP (SSP5) for this query by consulting its index IND (obtained by applying Decision Tree algorithms);

3. Each selected Super-Peer (SP6, SP8 and SP5) sends the query to relevant peers (P1, P3 and P4);

4. If there is no result in the index of SSP, then the SSP (SSP68) returns the query to the Super-Peer (SP6) to be treated with first step;

5. The final result of selected peers (P1, P3, P4 and P5) is returned (Index way + baseline way).

– The third step exploits the distributed knowledge Super-Super-Peers (SSP) network only without baseline approach. This step is very useful, for it allows us to see the use of the data mining in the P2P context and its effects on the performance of our proposed system. This step runs in three stages:

1. The Super-Peer (SP6) sends the query directly to its Super-Super-Peer (SSP68);

2. The Super-Super-Peer (SSP68) identifies (without the mapping) the relevant Super-Peers (SP8) that belong to this SSP (SSP68) and other SSP (SSP5) for this query by consulting its index IND (obtained by applying Decision Tree algorithms);

3. Each selected Super-Peer (SP6, SP8 and SP5) sends the query to relevant peers (P1, P3 and P4);

4. The final result of selected peers (P1, P3 and P4) is returned (Index way only).

## 7. SIMULATOR ARCHITECTURE

Since the beginning of P2P-applications, there has been a discussion over the usage of P2P-networks, whether they offer legal or illegal content. It is no secret that most of the shared data in common P2P-networks consists of illegal content. On the other hand**,** companies see the potential of P2P-technolgy. RedHat, i.e., distributes their Linux-Images over BitTorrent, because it would be too expensive to provide the necessary server-infrastructure including the server hardware and bandwidth capacity. And this is only one small example. There are many ways to use P2P-technology for legal content distribution; and P2P is still growing in popularity. Thus, it is important to improve new P2P technologies and make them more efficient. Unfortunately, it is not possible to test or simulate P2P-networks like a client-server architecture, where a complete overview over the destination network segment is always ensured. P2P networks are in a constant flow of peers connecting and disconnecting to the network. And this is just one example of the complexity of P2P. Due to these problems, it is nearly impossible to assess the performance of a new P2P application or algorithm without simulation. Accordingly, the need for suitable simulators for P2P-networks has evolved. There are several Peer-to-Peer simulators and which are presented herein:

P2PSim [25] is a discrete event simulator for structured overlay networks written in C++. It comes with seven Peer-to-Peer protocols implemented, including the more recent protocols Koorde [26] and Kademlia [27]. There are a number of different underlying network models, all of them; however, they are on a rather abstract level of detail, making it hard to simulate the dedicated overlay devices in the access networks mentioned above. P2PSim is largely undocumented and therefore hard to extend.

OverlayWeaver [28] is a Peer-to-Peer overlay construction toolkit written in Java which can be used for easy development and testing of new overlay protocols and applications. The toolkit contains a so-called Distributed Environment Emulator which invokes and hosts multiple instances of Java applications on a single computer; a fact that allows the simulation of up to 4,000 nodes. Since simulations have to be run in real-time and there is no statistical output, the toolkit's use as an overlay network simulator is very limited.

PlanetSim [29] is an object-oriented simulation framework for overlay networks and services written in Java. In addition to the overlay protocols Chord [30] and Symphony [31] there are several services like CAST and DHT available on application layer. PlanetSim offers only limited support to collect statistics and has a very simplified underlying network layer without consideration to bandwidth and latency costs. This makes it difficult to simulate heterogeneous access networks and terminal mobility. It is possible to visualize the overlay topology at the end of a simulation run, but there is no interactive GUI.

A more comprehensive survey of Peer-to-Peer network simulators can be found in [32], where the authors show that most available Peer-to-Peer network simulators have several major drawbacks, limiting them in use for research projects.

When implementing a P2P simulation, it is important to consider basic facts, including peer behavior, bandwidth, network topology because the principle behind locating the content usually affects the properties of the architecture, and; consequently, the simulation model as well as the different other aspects such as the dynamic changes in network size, peer capabilities, characteristics of the shared files, human behavior etc.

In order to understand how the simulators handle the different P2P-principles, it is important to take a closer look at how a P2P-networks works. In a network with one server and many clients (e.g. Windows Domain), the clients and the workload, which is consumed by the network, switches and routers can be controlled from the server. It is possible to refuse connections and have control over the network traffic and the data that flows through the network. With the server as the single point of failure, the complete network depends on the functionality of the server. Within a P2P network and its decentralized nature, every peer becomes a client and a server at the same time. Thus, the network is in a dynamic flow with clients connecting and disconnecting to the network. The network itself uses the physical structure of the internet to build its own virtual network. This is usually done by the P2P application. As a result, in a Point-to-Point connection, peers can transfer data through a virtual connection.

For our implementation and simulation, we used the Java programming language, the SimJava package. SimJava [24] is a process based on discrete event simulation package for Java, and on a discrete event simulation kernel. SimJava includes facilities for representing simulation objects as animated icons on the screen. A SimJava simulation is a collection of entities each running in its own thread. These entities are connected together by ports and can communicate with each other by sending and receiving event objects.

Our simulator is based on a set of tools such as WEKA that is a data mining platform (See Figure 10).

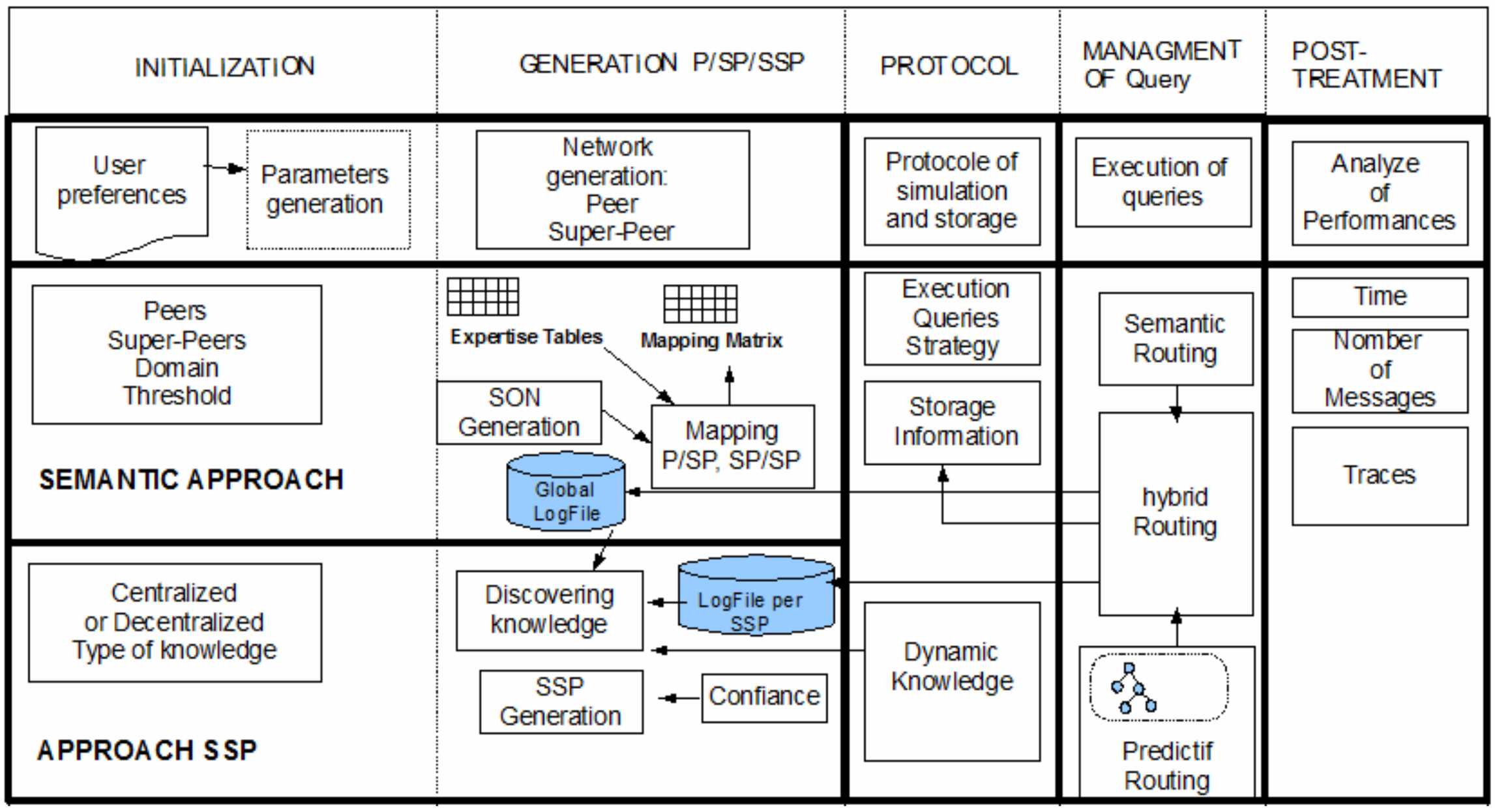

Figure 10. General Architecture of the Proposed Simulator

We have developed the necessary tools to interface the simulator with the various external components without any user intervention. This starts by initializing the parameters of the system, leading to the generation of a SON network with a level of Super-Peer (domain) which is juxtaposed to the peer's level. A third level domain-group complements the previous two levels; the trust (confidence) which is between the Super-Peers is needed to clarify the domain-group which is characterized by the knowledge using WEKA. The trust between two Super-Peers depends on the number of semantic links connecting them. The trust is useful where a Super-Peer SP leaves the network: peers attached to SP will then be attached to the Super-Peer with the highest degree of trust with SP. Then the simulation sending query begins and it is the characterizations of domain-groups that are used to route requests to the relevant Super-Peers. The aggregation of all results returned by each Super-Peer has processed the query, constituting the response to the submitted query. The generation of applications is ensured by peers. In fact, each peer P can generate a query by selecting elements of expertise that become components of the query Q. We say that a peer P is relevant to the query Q if the expertise of P contains at least a fraction of the components of Q. This is determined using the ability of a peer P to resolve a query Q.

So each peer generates a number N of queries that are derived from its expertise. After this phase generation of query, peers send their queries to their Super-Peers.

All queries exchanged within the network are stored in a file global LogFile. Thus, for a query Q, the file LogFile contains the following information: the identifier of the peer (P), which submitted the application, its Super-Peer (SP), the query (Q) itself, and the Super-Peer which responded favorably to this request.

## 8. RESULTS AND DISCUSSION

Decision Trees represent a supervised approach of classification. The WEKA classifier package has its own version of C4.5 known as J48. The Decision Tree algorithm can be summarized by these points:

1. Choose an attribute that best differentiates the output attribute values.

2. Create a separate tree branch for each value of the chosen attribute.

3. Divide the instances into subgroups so as to reflect the attribute values of the chosen node.

4. Terminate the attribute selection process, for each subgroup, if:

a- All members of a subgroup have the same value of the output attribute, terminate the attribute selection process for the current path, and label the branch on the current path with the specified value.

b- The subgroup contains a single node or no further distinguishing attributes that can be determined. As in (a), the branch with the output value seen by the majority of remaining instances is labeled.

5. Repeat the above process for each subgroup created in (3) that has not been labeled as terminal.

```
composanteW1 = k.f
|   composanteW2 = p.i: SP0 (26.0/11.0)
|   composanteW2 = f.p: SP0 (12.0)
|   composanteW2 = g.f: SP3 (26.0)
|   composanteW2 = g.h: SP0 (15.0)
composanteW1 = p.i: SP0 (50.0)
composanteW1 = f.l:  SSP17 (78.0/12.0)
composanteW1 = f.p:  SSP17 (159.0/14.0)
composanteW1 = d.o
|   composanteW4 = r.m: SP3 (38.0/16.0)
|   composanteW4 = i.c: SP3 (25.0)
|   composanteW4 = s.d:  SSP68 (28.0)
|   composanteW4 = h.i:  SSP59 (60.0/37.0)
|   composanteW4 = s.m:  SSP68 (36.0/6.0)
|   composanteW4 = d.m: SP3 (24.0)
composanteW1 = r.m:  SSP59 (393.0/138.0)
   ----
```

Figure 11. Example of a Decision Tree for SSP03

The models produced by Decision Trees are represented in the form of tree structures. A component of query indicates the class of the examples. The instances are classified by sorting them down the tree from the first component of the query to other component of the query. Decision Trees represent a supervised approach of classification. WEKA uses the J48 algorithm, which is WEKA'S implementation of C4.5 Decision Tree algorithm. J48 is actually a slight improved on the latest version of C4.5. It was the last public version of this family of algorithms before the commercial implementation C5.0 had been released. C4.5 was chosen for several reasons: it is a well-known classification algorithm; it has already been used in similar studies [33]; and, it can originate easily understandable rules. J48 is the Decision Tree classification algorithm. It builds a Decision Tree model by analyzing training data, and uses this model to classify user data. Figure 6 shows the results of running J48 Decision Tree algorithm.

Each line represents a node in the tree (See Figure 11). The second two lines, those that start with a '|', are child nodes of the first line. In the general case, a node with one or more '|' characters before the rule is a child node of the node that the rightmost line of '|' characters terminates at, if you follow it up the page. The next part of the line declares the rule. If the expression is true for a given instance, you either classify it if the rule is followed by a semicolon and a class designation–that designation becomes the classification of the rule–or, if it isn't followed by a semicolon, you continue to the next node in the tree (i.e. the first child node of the node you just evaluated the instance on). If the expression is instead false, you continue to the "sister" node of the node you just evaluated; that is, the node that has the same number of '|' characters before it and the same parent node.

Nodes that generate a classification, such as composanteW1 = j.m: SP1 (50.0), are followed by a number (sometimes two) in parentheses. The first number tells how many instances, in the training set, are correctly classified by this node, in this case 50 are. The second number, if it exists (if not, it is taken to be 0.0), represents the number of instances incorrectly classified by the node.

The classification of large datasets is an important data mining methodology. For our purposes, the most important figures here are the numbers of correctly and incorrectly classified instances. The output from the WEKA program is shown in Figure 11. In this output, the Decision Tree is able to classify approximately ninety two percent of the data correctly.

Each SSP operates with an index that keeps track of where contents concerning a query are located: when a SSP receives a query from a Super-Peer (in his group), it consults directly its index (See Figure 9), in order to determine:

1. In his group all Super-Peers (example SP0 and SP3) (or domains) are able to answer this query and

2. In other groups (example SSP68, SSP17...) (i.e. other SSP) all Super-Peers which are relevant to this query.

Evaluating the performance of P2P network is an important part to understand how useful it can be in the real world. As with all P2P applications, the first question is whether P2P is scalable. Our systems were evaluated with different set of parameters i.e. number of peers, number Super-Peer etc. Evaluation results were quite encouraging. There are many dimensions in which scalability can be evaluated: one important metric is the running time of a query. We run simulations on P2P network of different sizes. Each peer sends Query to its SP that in its turn sends the query to an SSP in order to find which Super-Peer(s) can answer the given query in the both architectures.

First, we modified the number of peers (300, 600,..., 5000 peers) and Super-Peers (10, 12 ,14, 16, 20,..., 54) in both architectures to measure the execution time. Second, the most popular measure for the effectiveness of our systems is precision.

Precision = (relevant answered peers in architecture (DK-Bis))/(relevant answered peers in architecture-baseline).

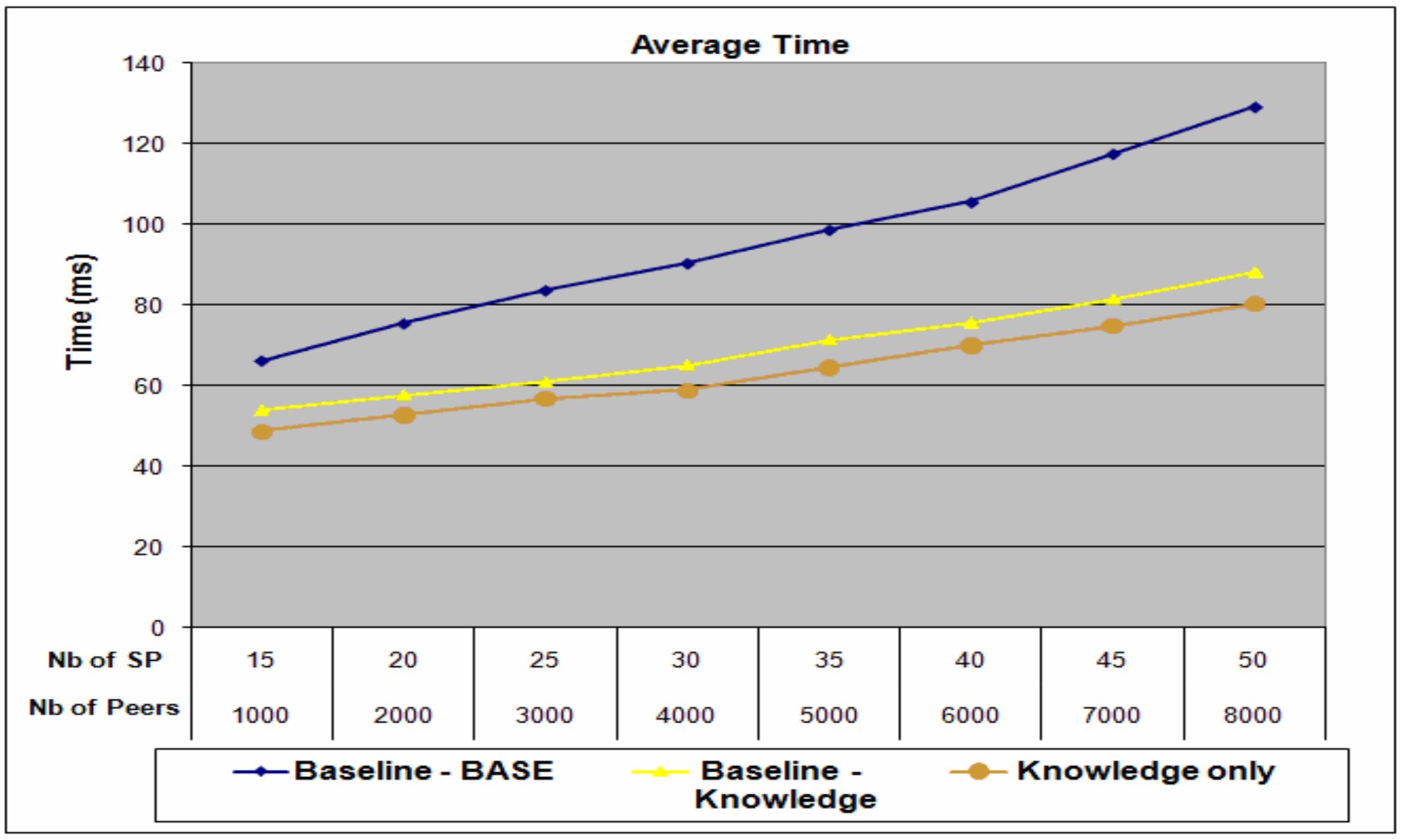

Figure 12. Execution Time

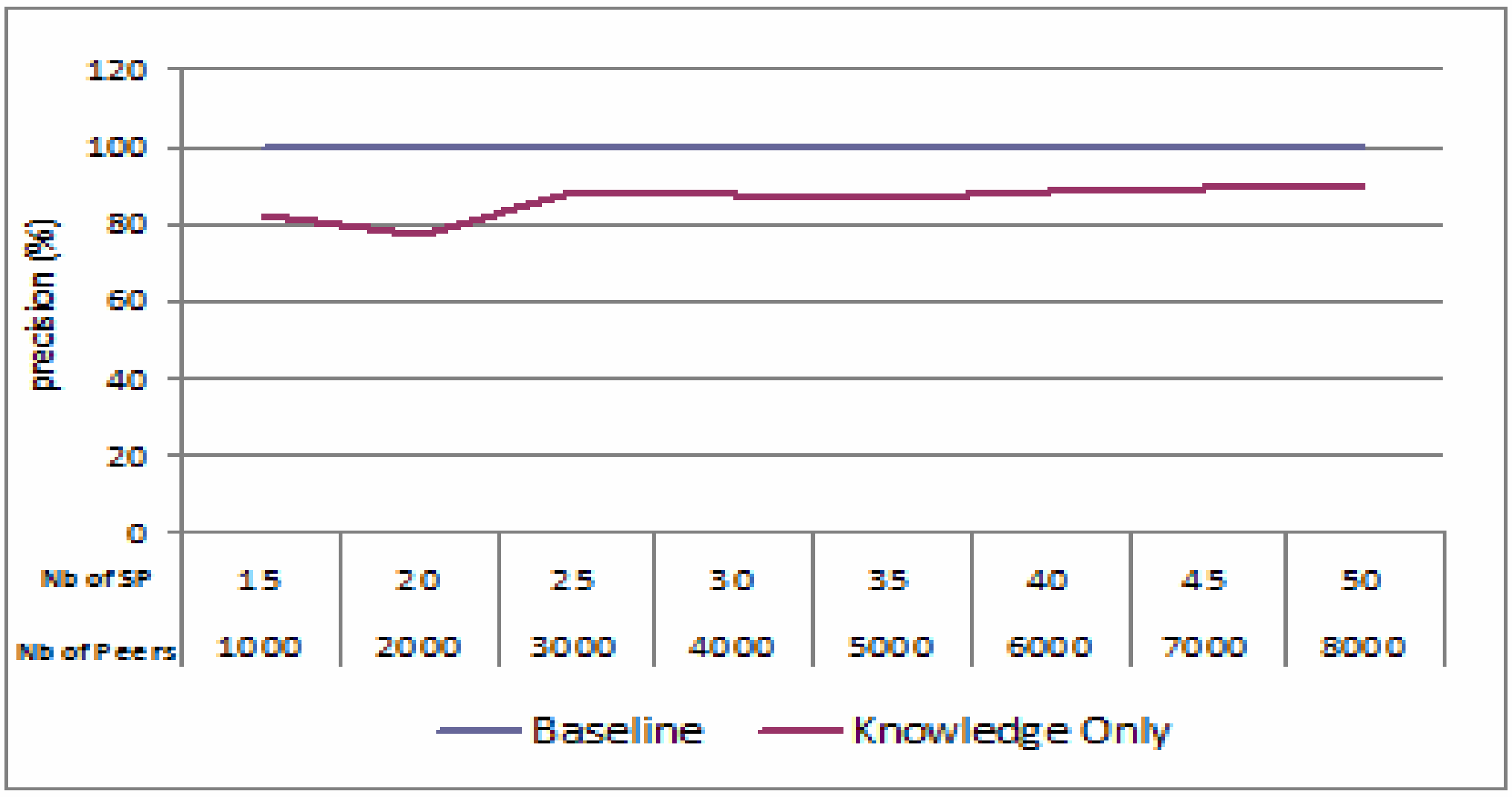

Figure 13. Precicion Rate

The Graphs shown in figures 12 and 13 are the results of our simulations. They demonstrate the performance of using the Super-Super-Peer with a Decision Tree for routing queries to relevant P2P domains (SP). In the first observation, the difference in the execution times between 1000 and 2000 peers in the architectures DK and DK-bis is small (See Figure 10). Measurements shown in Figure 12, show that the answering time in Architecture with knowledge is less about 35 % than the answering time in Architecture-baseline, this is due to the processing of queries using Decision Tree that we have proposed. Therefore, this shows the scalability of our architecture. Measurements in Figure 13 show the accuracy (precision) of the architecture with knowledge only compared to the baseline architecture. We could observe that there is almost a linear line in precision between these architectures, which reflects the stability of our architecture with the increasing number of peers and Super-Peers.

Finally, our Prototype in grouping P2P domains (P2P) raises some interesting performance issues. We perform experiments to demonstrate how the presence of grouping domains affects their performance, in addition to illustrate how grouping domains can improve the scalability of the overall system.

## 9. CONCLUSIONS

In this paper, we proposed an architecture using distributed classification for P2P networks. We captured the traffic of queries and their results in the baseline architecture, preprocessed and labeled the data, and built several models using a combination of different attributes in the training-data set. We observed that the accuracy of the classifier increases significantly when we take a bigger size of the network in the baseline architecture. This implies that the accuracy of the classifier increases when we get more information about the domain of peers. To detect domains of peers, the decision-tree algorithm (J48) needs to be implemented in an added level called SSP.

One important area for improvement is performance. Some of the options for improving performance were discussed in the evaluation of P2P Network and include: improvements in the answering time of a given query. By the analysis of the outcome of the experiments, we demonstrated that the integration of the data mining in the P2P context of our proposed system has gave a high performance and therefore it is scalable.

## Authors

Dr. Anis Ismail, Born in Lebanon, April 1979, works as system and network administrator and instructor at the Lebanese University, University Institute of technology, Sidon, Lebanon. He has a B.S. degree in Telecommunication and Networking Engineering from the Lebanese University (LU), an M.S. in Computer Science from the American University of Science and Technology (AUST) in Lebanon, and a Ph.D. in Computer Science from the University of AIX-Marseille, France.

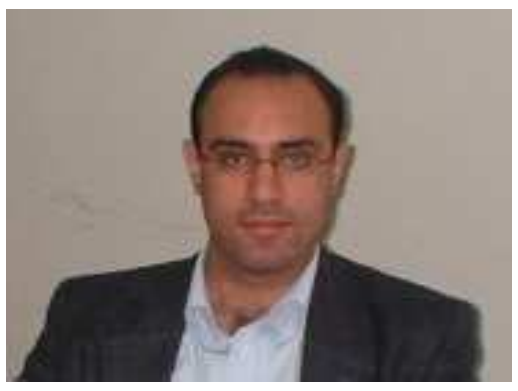

Dr. Aziz M. Barbar is the Chairperson of the Department of Computer Science at the American University of Science & Technology (AUST), Lebanon. He has a Ph.D. in Computer Science from the University of Nice-Sophia Antipolis (France). His research interests include Database Reverse Engineering, Data Mining and Natural Language Processing. Dr. Barbar is currently the Vice-President of the Lebanese Information Technology Association (LITA), and the Chair of the IEEE Computer Chapter, Lebanon.

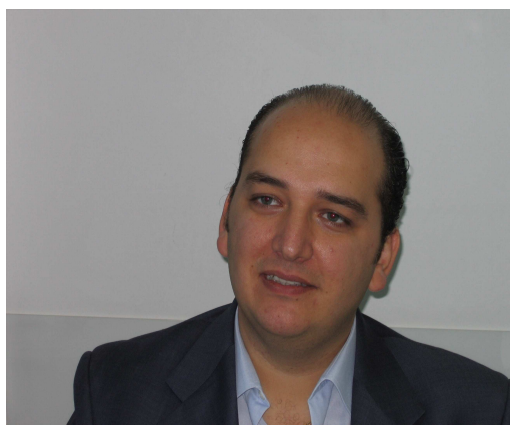